\documentclass[preprint,showpacs,preprintnumbers,amsmath,amssymb,aip]{revtex4-1}

\usepackage{graphicx}
\usepackage{dcolumn}
\usepackage{bm}
\usepackage{here}
\usepackage{float}

\begin{document}

\title{Detection of radiation torque exerted on an alkali-metal vapor cell}

\author{Atsushi Hatakeyama}
\email[]{hatakeya@cc.tuat.ac.jp}
\affiliation{Department of Applied Physics, Tokyo University of Agriculture and Technology, Koganei, Tokyo 184-8588, Japan}
\author{Runa Yasuda}
\affiliation{Department of Applied Physics, Tokyo University of Agriculture and Technology, Koganei, Tokyo 184-8588, Japan}
\author{Yutaka Goto}
\affiliation{Department of Applied Physics, Tokyo University of Agriculture and Technology, Koganei, Tokyo 184-8588, Japan}
\author{Natsumi Chikakiyo}
\affiliation{Department of Applied Physics, Tokyo University of Agriculture and Technology, Koganei, Tokyo 184-8588, Japan}
\author{Takahiro Kuroda}
\affiliation{Department of Applied Physics, Tokyo University of Agriculture and Technology, Koganei, Tokyo 184-8588, Japan}
\author{Yugo Nagata}
\affiliation{Department of Physics, Tokyo University of Science, Kagurazaka, Shinjuku-ku, Tokyo 162-8601, Japan}

\date{\today}

\begin{abstract}
We have developed a torsion balance to detect the rotation of a cell containing spin-polarized gaseous atoms to study angular momentum transfer from gaseous atoms to solid. A cesium vapor cell was hung from a thin wire in a vacuum chamber, and irradiated from the bottom with circularly polarized light tuned to the $D_2$ transition to polarize cesium atoms in the cell. By varying the light helicity at the resonance frequency of the torsion balance, we induced forced rotational oscillation of the cell and detected radiation torque exerted on the cesium vapor cell through the cesium atoms inside. The torque was particularly large when both hyperfine levels of cesium atoms were optically pumped with application of a longitudinal magnetic field. Further detailed study will provide new insights into spin-transfer processes at the gas-solid interface.
\end{abstract}


\maketitle

\section{INTRODUCTION}
Circularly polarized light transports angular momentum and exerts torque on an irradiated object. In 1936, Beth demonstrated that a hung half waveplate rotated when circularly polarized light passed through it and flipped the light helicity.\cite{Beth36}
From a quantum viewpoint, each photon in left (right) circularly polarized light has a spin component of $\hbar$ ($-\hbar$) in the direction of light propagation (e.g., Refs.\cite{Jak98, Hec98}), and transfer of the photon spins to an object leads to positive (negative) radiation torque on the object.
Spin-dependent radiation forces are interesting topics in recent optomechanics researches.\cite{Mag18}

With regard to spin transfer from photons to materials, photon spins can be effectively transferred and stored as atomic spins in gaseous atoms by optical pumping.\cite{Hap72} The spin-polarized atoms fly to and collide with the walls of a gas container (often called a cell). The spin polarization relaxes almost completely following a single collision with surfaces, such as glass and metal.\cite{Sek18} Some coating materials, such as paraffin, suppress spin relaxation to a great extent,\cite{Bud13} but the polarized atomic gas eventually loses its spin angular momenta. Although spin relaxation mechanisms at surfaces are not completely understood for all types of surfaces,\cite{Bou66} this spin relaxation is considered to involve the transfer of spin angular momenta from the gaseous atoms to the cell, from the viewpoint of the conservation of angular momentum. Under specific conditions, the spin polarization of alkali-atom nuclei in alkali-metal hydrides was achieved by spin transfer from an optically pumped alkali-metal vapor.\cite{Ish07} There are various types of spin relaxation and transfer processes, but any type of spin transfer must rotate the cell macroscopically. 

Here, we have developed a torsion balance to detect the macroscopic rotation of a cell containing spin-polarized gaseous atoms, to study angular momentum transfer from gaseous atoms to solid. Dilute cesium (Cs) atoms in a vapor contained in an evacuated cell hung from a thin wire were spin-polarized by absorbing circularly polarized light and undergoing spontaneous emission repeatedly during optical pumping. The spin polarization was maintained under a uniform longitudinal magnetic field until the atoms collided with the cell surface, where they lost their spin polarization. We varied the helicity of the circularly polarized light at the resonance frequency of the torsion balance and clearly showed that optical pumping of Cs atoms exerted torque on the cell. The induced rotational oscillation was enhanced when the atomic gas achieved high polarization under appropriate conditions of light polarization, light frequency, longitudinal magnetic field, and atomic density of the gas. Further quantitative and detailed studies will hopefully reveal spin-transfer processes at the gas-solid interface from a mechanical viewpoint.
We note that kinetic effects depending on atomic internal states to study the interactions of gaseous atoms with surfaces were also discussed previously.\cite{GHI83}

\section{EXPERIMENTAL APPARATUS}

\begin{figure}
\includegraphics[width=8cm]{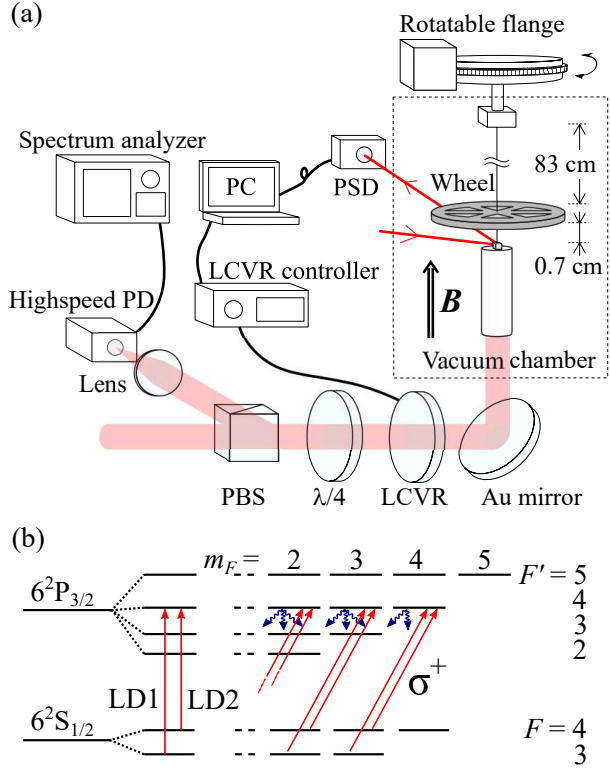}
\caption{\label{Fig1}(a) Schematic of the experimental setup. The cell was hung together with the wheel in the vacuum chamber under a longitudinal magnetic field ${\bf B}$ and irradiated from the bottom with the laser beam. PBS: polarizing beam splitter; $\lambda/4$: quarter waveplate; LCVR: liquid crystal variable retarder; PSD: position sensitive detector; PD: photodiode. (b) Relevant energy levels of the cesium (Cs) atom. $F$ and $F'$ are the total angular momenta of the ground and excited states, respectively, and $m_F$ represents their magnetic quantum numbers. The tuning of two lasers, LD1 and LD2, is indicated by red arrows. Optical pumping cycles are also illustrated in the case of the $\sigma^+$ pumping light  (red arrows: excitation by absorption; blue wavy arrows: deexcitation by spontaneous emission).}
\end{figure}

Figure 1 shows a schematic representation of the experimental setup. It is basically a torsion balance made up of a gas-containing cell hung with a thin wire in a vacuum. The cell contained a Cs vapor in equilibrium with a small amount of Cs metal. 
The cell was sealed off under high vacuum conditions and Cs atoms flew freely without collisions between the cell walls.
Cs metal has a relatively high melting temperature of $28^{\circ}$C, and Cs atomic densities in the vapor are $6\times 10^{10}$~cm$^{-3}$ at $30^{\circ}$C and $5\times 10^{11}$~cm$^{-3}$ at $50^{\circ}$C.\cite{Nes63} The ground state of a Cs atom has a 1/2 electron spin and a 7/2 nuclear spin; there is no orbital angular momentum. The ground state splits into two hyperfine levels labeled as total angular momenta $F=4$ and $3$, respectively. Cs atoms thus store up to $4\hbar$ angular momentum in the ground state.

The Cs vapor cell was cylindrical, 20~mm in outer diameter and 63~mm in outer length, with a short (5 mm), pinched-off stem. The cell was made of quartz glass. The glass wall was 0.5~mm thick to reduce the weight and inertia momentum of the cell, which were $5$~g and $5 \times 10^{-7}$~kg m$^2$, respectively. 

The cell was hung with a short tungsten (W) wire 10~$\mu$m in diameter and 0.7~cm in length. The wire was connected to a wheel hung with a long W wire 10~$\mu$m in diameter and 83~cm in length. The long wire was connected to the top flange of the vacuum chamber. The wheel had a relatively large mass and a large moment of inertia; $17$~g and $7 \times 10^{-6}$~kg m$^2$, respectively.
The longer wire and larger momentum of inertia of the wheel resulted in a lower resonance frequency of rotational oscillation of the wheel than the cell attached to the short wire. Thus, the wheel acted to isolate vibration from noise input through the top flange of the chamber.\cite{Sau17}
The wheel was made of black-anodized aluminum. We avoided using magnetic materials even for small screws contained in the wheel, to reduce the magnetic influence from the outside. The wheel could be made of dielectric material, which may be problematic when charged up, but might further reduce the magnetic influence of eddy currents.

The top flange holding the long wire was electrically rotatable, and was used to damp the rotational oscillation of the wheel; otherwise it would have kept oscillating out of the range of the detector ($30$~mrad in the current setup; see below) even for 1 week in a vacuum chamber. Once the rotational oscillation entered the detector range by manual control of the rotatable flange, it was further damped by a PC-based servo system. This servo system enabled rapid recovery from strong external disturbances, such as earthquakes. We turned off the servo and allowed the cell to oscillate freely during measurements.

The vacuum chamber made of stainless steel was evacuated with an ion pump down to $1\times 10^{-5}$~Pa and placed on an optical table (table top dimensions: $2.4$~m$\times 1.5$~m$\times 0.3$~m) located on the 4th floor of a building. The table top was supported with four air-cushioned legs, although we released air from the legs during measurements to reduce external vibration noise.
The main body of the chamber enclosing the cell and the wheel was cylindrical, with an inner diameter of 150~mm, and had some ports. A relatively narrow glass tube (52~mm in inner diameter), as a vacuum chamber surrounding the cell, caused a problematic electrostatic interaction between the cell and the tube; therefore, a metal one was used.
The chamber was equipped with electric heaters to control the temperature of the cell. The bottom flange was set hotter than the main body of the chamber to prevent vaporized Cs metal from condensing and forming a film on the bottom window of the cell.

A 5-mm glass cube, the side surfaces of which were coated with aluminum, was attached with optical glue to the top window of the cell and also connected to the short W wire. A probe laser beam was reflected on the cube to measure the rotation of the cell. The position of the reflected beam spot was measured with a position-sensitive detector (PSD), which had a 12-mm wide aperture and a sub-micrometer resolution. The PSD was placed 20~cm from the cell center.

Two distributed feedback laser diodes (DFB LDs) running at Cs $D_2$ transitions at a wavelength of 852~nm were used to polarize Cs atoms in the cell. One DFB LD (designated as LD1) was tuned to the excitation from the ground state hyperfine level $F=3$, while the other (designated as LD2) was tuned to the $F=4$ level. Two laser beams from the two DFB LDs were injected into a single-mode polarization-maintaining fiber with the same linear polarization, and the output beam from the fiber, with a power ratio of about 1 to 2 for LD1 to LD2, was then injected into a tapered amplifier. An amplified pumping beam passed through a single-mode fiber to make its spatial mode Gaussian TEM$_{00}$. The polarization of the pumping beam was controlled with a quarter waveplate and a liquid crystal variable retarder (LCVR) before the beam was reflected into the cell with a gold mirror. We checked the polarization of the laser beam just before it entered the vacuum chamber by tentatively inserting a quarter waveplate and a polarizing beam splitter. The final beam diameter ($1/e^2$) and power were 8~mm and 50~mW, respectively.

The circularly polarized pumping beam resulted in spin polarization of Cs atoms in the ground state by a standard optical pumping technique. The helicity of the polarization, changeable with the LCVR, determined the direction of atomic spins. The optical density of Cs atomic vapor for $D_2$ resonant light was high. Using an independent setup, we confirmed that the transmission of the pumping light through an identical Cs vapor cell was less than 1\% at $50^{\circ}$C. Thus, almost all photons in the pumping beam were absorbed and emitted by Cs atoms. 

High polarization was achieved by exciting Cs atoms from both ground state hyperfine levels. The ground state hyperfine levels were separated by 9.19~GHz, while the hyperfine levels of the upper state were unresolved due to Doppler broadening of the $D_2$ line. Precise frequency tuning of two DFB LDs was achieved by referring to Doppler-free spectra obtained by saturation absorption spectroscopy. We also monitored the beat frequency of two DFB LDs continuously with a high-speed photodiode (PD) and a spectrum analyzer, and adjusted during measurements to confirm that the frequency difference matched the ground state hyperfine splitting.

High polarization is established by several cycles of laser excitation and spontaneous emission,\cite{Hap87} occurring within a lifetime of 30~ns. Established polarization was maintained with a uniform longitudinal magnetic field; we applied a 0.2~mT field in this experiment. In the absence of a longitudinal field, the polarized spins would precess at a rate of 3.5~MHz/mT around a stray transverse field and degrade polarization. With application of a longitudinal field, polarized atoms maintained their polarization until flying without collisions with other gas molecules to the cell walls; the typical time of flight was 50~$\mu$s. The Cs atoms lost their polarization almost completely following a single collision with the quartz surface of the cell.\cite{Sek18}

\section{EXPERIMENTAL RESULTS}

Figure 2 (a) shows the typical evolution of rotation angle of the cell freely oscillating over $5\times 10^4$~s with no pumping light and application of a longitudinal magnetic field at room temperature. The most notable rotational oscillation had a period of 1,300~s. This oscillation was caused by the wheel from which the cell was hung with the short W wire. This and other oscillation frequency components were manifested in the Fourier transform spectrum shown in Fig. 2 (b). The lowest frequency peak at 0.77~mHz corresponds to the rotational oscillation of the wheel. The highest peak at 0.46~Hz was caused by lateral oscillation of the wheel hung with the 83~cm W wire from the chamber flange. A higher component originating from lateral oscillation of the cell hung with the 0.7~cm wire was out of the bandwidth of the measurement with a sampling rate of 1 s. The frequency peak at 29~mHz corresponded to the rotational oscillation of the cell hung from the wheel with the 0.7~cm wire.
We confirmed that these resonance frequencies reasonably agreed with estimations made from the moments of inertia for the wheel and the cell, as well as the lengths and the Young's modulus of the W wires.

The inset of Fig.~2 (b) shows a magnified resonance peak at 29~mHz, which was the focus of the following measurements. From fitting with a Lorentz function, we inferred the Q-factor of this torsion balance to be about 1,000.  

\begin{figure}
\includegraphics[width=8cm]{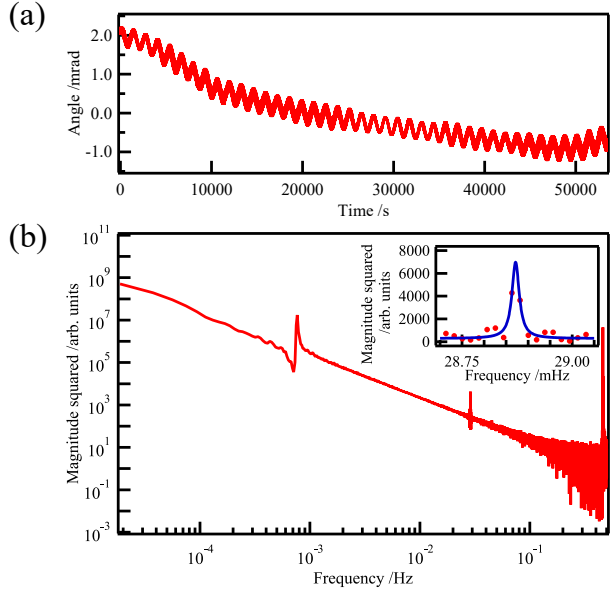}
\caption{\label{Fig2}(a) Time evolution of the rotation angle of the freely oscillating cell and (b) its Fourier transform spectrum. Inset: resonance peak at 29~mHz for the cell and its Lorentzian fitting.}
\end{figure}

To investigate cell rotation induced by spin-polarized Cs atoms, we employed a forced oscillation method at a resonance frequency of 29~mHz. We changed the helicity of the pumping light continuously between right- and left-circular polarization with the LCVR at 29~mHz, and measured the amplitude of the resulting oscillation at the same frequency. The cell temperature was 50$^{\circ}$C. 
At this temperature, we estimated that the transport rate of atomic spins to the cell walls was roughly balanced with the impinging rate of photon spins if one atom carried 2$\hbar$ angular momentum in average.
Figure 3 shows the typical time evolution of the oscillation amplitude under several conditions. LD1 and LD2 that produced the pumping light were tuned to the $F=3 \to F'=4$ transition and $F=4 \to F'=4$ transition, respectively, with the frequency difference equal to the Cs hyperfine splitting 9.19~GHz. The pumping light was incident to the cell before the measurement started. The amplitude measured before period (a) was a typical background level ($\sim10~\mu$rad). During period (a), the LCVR switched the helicity of the circularly polarized light, and then the amplitude of the forced oscillation increased. During period (b), the light polarization was changed with the quarter waveplate and the polarization was switched between two orthogonal linear polarizations. No spin polarization was thought to be produced in the Cs atomic gas with this linear polarized light, and the forced oscillation decayed. 
The residual amplitude was due probably to radiation pressure effects as also discussed later.
During period (c), we switched the light polarization back to circular, and observed recovery of the oscillation. During period (d), the longitudinal magnetic field was set to zero from 0.2~mT. Under these conditions, polarization of Cs atoms was not maintained due to the residual transverse magnetic field, and the forced oscillation was damped. 
In this case, it can be considered that, although angular momenta were once transferred from photons to gaseous atoms, the atoms released their angular momenta not to the cell but to environment that produced the transverse magnetic field.

\begin{figure}
\includegraphics[width=8cm]{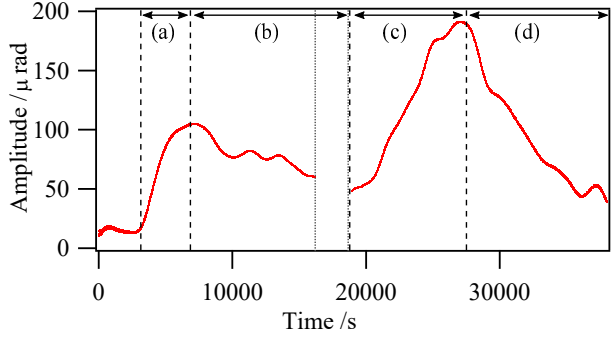}
\caption{\label{Fig3}Time evolution of the amplitude of forced rotational oscillation of the cell under different optical conditions (a)-(d): (a) and (c) optimum optical pumping; (b) orthogonal linear light polarizations; (d) no longitudinal magnetic field. The measurement of rotational oscillation was interrupted in the end of the period (b). }
\end{figure}

The above results indicated that radiation exerted torque on the cell through polarized Cs atoms within the cell. We performed further measurements under different conditions. Figure 4 summarizes the amplitudes of the steady-states of forced oscillation under various conditions for light frequency tuning, light polarization, longitudinal field, and atomic density. Due to uncontrollable perturbations in the moment, the obtained results had uncertainty of about 50\%. Condition A was expected to achieve the highest polarization of Cs atoms as in periods (a) and (c) in Fig.~3, and actually showed the largest forced oscillation. Other conditions from B to E were expected to show low polarization, and small amplitudes of forced oscillation were actually obtained. Under condition B, polarization of the pump light was linear. The longitudinal magnetic field was turned off under condition C. When we set the laser frequencies completely out of the Doppler-broadened lines in condition D, there was no interaction between the pumping light and Cs atoms in gas phase. It should be noted that a small oscillation was forced even under condition D. We considered that some force exerted directly on the cell from pumping light induced this rotational oscillation. Finally, under condition E, the temperature of the cell was decreased to reduce Cs atomic density by a factor of 10. 
At this density, the pumping light was still absorbed by 80\%, but we consider that the magnitude of angular momentum transferred to the cell through the atoms was reduced according to the atomic density.
These conditions resulted in less transfer of angular momenta from photons to the solid. 

Taken together, the above results suggest that the macroscopic rotation of the cell was induced by radiation-induced torque through polarized Cs atomic gas within the cell. The extent to which the spin transfer mechanism at the cell surface contributed to the observed forced oscillation is not yet clear. As seen in the measurements under condition D, there were effects that were unrelated to optical absorption by Cs atoms, presumably radiation pressure effects.\cite{Beth36,Mar84} We roughly estimated the expected amplitudes of the oscillation from the radiation torque originating from photon spins ($2\times 10^{-17}$~N m) and the Q-factor of the torsion balance ($\sim 1,000$), and obtained amplitudes $\sim$ 2~$\mu$rad, which was smaller than the observed amplitudes. 
Further detailed studies, such as dependences on the pumping power, atomic density, and the degree of spin polarization, including radiation trapping effects,\cite{Tup87} are required to quantitatively evaluate the process of spin transfer from atoms to solids.

\begin{figure}
\includegraphics[width=8cm]{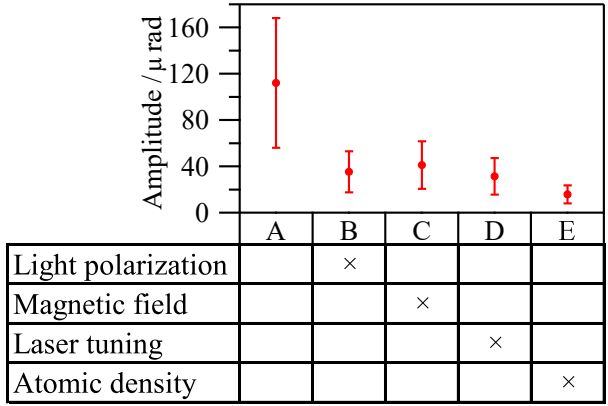}
\caption{\label{Fig4}Steady-state amplitudes of forced rotational oscillation of the cell under various optical pumping conditions. The error bars, roughly estimated to be 50\% from repeated measurements, are also displayed. The bottom table indicates parameters intentionally set to inappropriate values to achieve high spin polarization.}
\end{figure}

\section{CONCLUSIONS}
In conclusion, we have developed a torsion balance to detect the macroscopic rotation induced by spin transfer from gaseous atoms to solid. We demonstrated the radiation torque exerted on a Cs vapor cell through Cs atoms within the cell from circularly polarized light resonant to the Cs transitions. Prominent rotation of the cell was observed under conditions where atoms were highly polarized; inappropriate light frequencies and polarizations, a lack of longitudinal magnetic fields, and low atomic density suppressed the amplitude of the forced oscillation. Further quantitative studies will determine the extent to which the observed torque originated from the transfer of angular momenta from optically polarized atoms to the cell through collisions with the surface. Our mechanical method will provide a new means of investigating angular momentum transfer processes at the interfaces between gases and solids. 

This study can be regarded as a combination of the Einstein-de Haas (EdH)\cite{Ein15} and Beth experiments.\cite{Beth36} In the EdH experiment, the rotation of a magnetic material was shown to be induced by flipping its magnetization. This result, obtained before the development of quantum mechanics, demonstrated that magnetization was related to angular momentum. It can now be explained as compensation of the change in direction of atomic spins in a solid, a source of magnetization, by the rotation of the solid to fulfill the conservation of angular momentum. 
In our experiment, with regard to angular momentum transfer from photons to matter, microscopic processes can be well understood as optical pumping, and were much more controllable than in the Beth experiment because atom-photon interactions can be precisely adjusted. With regard to angular momentum transfer from atomic spin to solid rotation, the experiment described here allows control of a wide variety of parameters; spins can be manipulated both optically and magnetically during the flight of atoms to the cell surface, the properties of which can also be modified using surface coating techniques. 

\section*{ACKNOWLEDGMENTS}
We thank K. Harii, H. Chudo, M. Ono, and E. Saitoh for fruitful discussions. This work was supported by a Grant-in-Aid for Scientific Research on Innovative Area, ``Nano Spin Conversion Science'' (Grant Nos. JP15H01013 and JP17H05178) from MEXT, Japan.


%

\end{document}